# Flexibility of brain regions during working memory curtails cognitive consequences to lack of sleep

**Nina Lauharatanahirun[1,2,6,7*], Kanika Bansal[1,3], Steven M. Thurman[1], Jean M. Vettel[1,4], Barry Giesbrecht[4,8], Scott Grafton[4,8], James C. Elliott[4,8], Erin Flynn-Evans[5], Emily Falk[2] & Javier O. Garcia[1*]**

[1]Human Research and Engineering Directorate, US Combat Capabilities Development Command Army Research Laboratory, Aberdeen Proving Ground, MD
[2]Annenberg School for Communication, University of Pennsylvania, Philadelphia, PA
[3]Department of Biomedical Engineering, Columbia University, New York, NY
[4]Psychology and Brain Sciences, University of California, Santa Barbara, Santa Barbara, CA
[5]National Aeronautics and Space Administration Ames Research Center, Mountain View, CA
[6]Department of Biomedical Engineering, Pennsylvania State University, State College, PA
[7]Department of Biobehavioral Health, Pennsylvania State University, State College, PA
[8]Institute for Collaborative Biotechnologies, University of California, Santa Barbara, Santa Barbara, CA
*Correspondence to nina.lauhara@gmail.com or javiomargarcia@gmail.com

## Abstract

Previous research has shown a clear relationship between sleep and memory, examining the impact of sleep deprivation on key cognitive processes over very short durations or in special populations. Here, we show, in a longitudinal 16 week study, that naturalistic, unfettered sleep modulations in healthy adults have significant impacts on the brain. Using a dynamic networks approach combined with hierarchical statistical modelling, we show that the flexibility of particular brain regions that span a large network including regions in occipital, temporal, and frontal cortex increased when participants performed a working memory task following low sleep episodes. Critically, performance itself did not change as a function of sleep, implying adaptability in brain networks to compensate for having a poor night's sleep by recruiting the necessary resources to complete the task. We further explore whether this compensatory effect is driven by a (i) increase in the recruitment of network resources over time and/or (ii) an expansion of the network itself. Our results add to the literature linking sleep and memory, provide an analytical framework in which to investigate compensatory modulations in the brain, and highlight the brain's resilience to day-to-day fluctuations of external pressures to performance.

# Introduction

Sleep and memory are closely intertwined (Stickgold, 2005; Walker & Stickgold, 2004, 2006). While converging evidence from molecular (Graves, Pack & Abel, 2001) to behavioral (Hennevin et al., 1995) disciplines provide the foundation for the known relationship between sleep and memory, the mechanistic underpinnings and the strength of these relationships is still an area of active debate (Chee & Chuah, 2008; Donlea, 2019; Lowe et al., 2017; Manoach & Stickgold, 2019; Marshall et al., 2020; Stickgold & Walker, 2005). In one body of research, highly controlled laboratory experimentation where animals and humans' sleep schedules were unnaturally modified elicited failures in memory formation and/or retrieval (Walker & Stickgold, 2004). Similarly, in the cognitive domain, working memory performance is often reduced as a result of sleep deprivation (Habeck et al., 2004; Lim & Dinges, 2010) and linked to specific networks in the brain (Chee & Choo, 2004; Luber et al., 2008). While these scientifically rigorous laboratory studies have provided foundational knowledge and have led to well-established theories of sleep effects on behavior and the brain, researchers are now beginning to test the robustness of sleep effects on memory and extend experimentation to more naturalistic contexts providing ecologically valid results (Matusz et al., 2018) that may better capture the influence of sleep on working memory.

To expand our understanding of this sleep-memory relationship, we investigate the working memory consequences to unconstrained, naturalistic fluctuations in sleep using a standard *visual working memory* (VWM) task. This task is thought to probe foundational elements of working memory and task execution (D'Esposito & Postle 2015) like attentional filtering and capacity (Drummond et al., 2012), elements that severely limit more downstream cognitive functions like language comprehension (Baddeley, 2003), learning (Mayer and Moreno, 1998) and decision making (Luck and Vogel, 2013). Even more generally, working memory is crucial to simultaneously coordinate diverse cognitive processes when multiple goals are active. Evidence of this coordination in the brain comes from studies that borrow methods from network science where *dynamic community detection* has been used to investigate temporal changes in networks (Mucha et al., 2010; Garcia et al., 2018), characterizing the emergence and reconfiguration of functional networks across time. Previous research employing this technique has demonstrated that memory performance depends on the brain's frontal executive center to 'flexibly' reconfigure under task demands (Braun et al., 2015). However, if and how this *network flexibility* – the probability that a brain region alters its functional affiliation across time – compensates for changing physiological states remains unexplored.

Here, we employ dynamic community detection to investigate the impact of sleep on working memory within the human brain. In particular, we examine how naturalistic changes across several weeks in sleep duration (total sleep time) affect the *resilience* or *sensitivity* of brain dynamics and its relation with working memory performance. In order to investigate this relationship and account for the longitudinal and individual-specific effects of network metrics and sleep on behavior, we used a multilevel modeling approach where we model the unique and dependent contributions of dynamic network properties of the brain and total sleep time (TST) of the previous night, to behavioral performance in a working memory task. We use this powerful analytical approach as it provides a principled way to find theoretically driven multivariate associations between brain dynamics and behavioral performance (Aarts et al., 2014; Harrison et al., 2018). We find that flexibility of regions in visual, frontal, and temporal cortex is modulated by naturalistic fluctuations in the total sleep time across individuals. Given that this flexibility is related to behavioral performance, modulation by sleep suggests a unique resilience to external pressures, like fluctuations in sleep, that is possible due to our brain's ability to dynamically reconfigure in the face of physiological state changes.

# Results

Forty-two subjects participated in a longitudinal study where their sleep was continuously monitored for 2 to 16 weeks with actigraphy and sleep logs (Thurman et al., 2018). Every 2 weeks, subjects participated in an fMRI visual working memory task, among other tasks (see Thurman et al., 2018). After standard pre-processing (see *Methods)* of the fMR images, time courses of the blood oxygen level-dependent (BOLD) response were extracted from regions corresponding to the Desikan-Killiany atlas parcellation (Desikan et al., 2006). In the primary analysis, for each session and individual, we extracted time-varying functional connectivity patterns via a windowed wavelet coherence approach (Grinsted et al., 2004). These connectivity patterns were then subjected to a *dynamic community detection* analysis (Mucha et al. 2010, Garcia et al., 2018) that distilled the complex connectivity patterns into clusters of brain regions. Previous research has shown that the flexibility of sub-networks in engaging with each other is essential for efficient execution of cognitive function (Bassett et al. 2011, Braun et al. 2015). Therefore, we focused our analysis on the average 'flexibility' of each brain region by computing its tendency to change functional affiliation across time.

## Sleep, performance, and network flexibility.

Sleep from the night prior to each experimental session was estimated via an actigraph watch (Thurman et al., 2018). As displayed in Figure 1A, despite no manipulation of sleep schedules, substantial variability in average sleep across the multiple sessions was observed for each subject, where the average night's sleep prior to testing was 451.63 minutes, i.e., about 7.5 hours, (SD = 107.9). Visual working memory performance was determined by the number of correct targets identified (Fig. 1B; see *Methods* for details).

After deploying the *dynamic community detection* methodological approach on time-evolving patterns of functional connectivity, the flexibility of each brain region (i.e., node) was estimated, which captures the reconfiguration of functional networks across time and has been demonstrated to be linked to working memory (Braun et al. 2015). Across sessions and subjects, the mean flexibility of the brain regions on a scale from 0.36 – 0.88 was 0.62 (SD=.06; Figure 1C), indicating substantial change in community affiliations across the approximately 5 minutes of subjects performing the working memory task. Inspection of the spatial distribution of flexibility (Figure 1D) shows three clusters of particularly high flexibility in the medial portions of cortex in temporal and frontal lobes that included bilateral parahippocampal cortex (left, M = 0.71, right M = 0.69), entorhinal regions (left, M = 0.69, right, M = 0.70), and regions including the temporal pole (left, M = 0.69, right, M = 0.70).

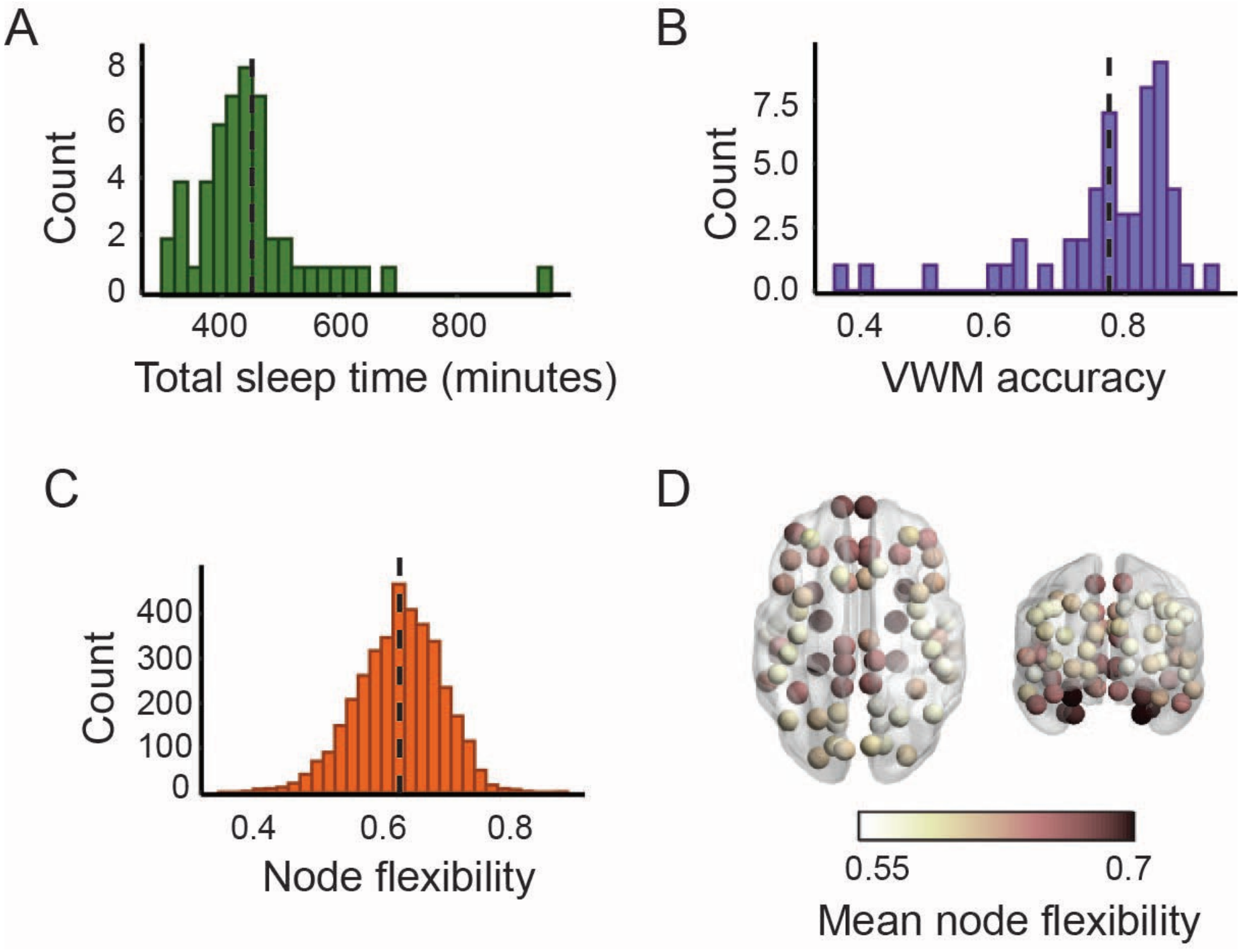

**Figure 1: Histograms and means of data types. (**A) Histogram displaying the frequency of estimated total sleep time (TST) across participants and subjects. (B) Histogram displaying the frequency of VWM performance, as estimated by the percentage of targets identified within a session. Flexibility is visually shown as a frequency histogram for all nodal estimates (C), and average flexibility within each node is displayed on the inset brain image (D) where each orb is plotted at the centroid of the region and colored by it’s mean flexibility. Mean TST, performance, and flexibility are all indicated with a vertical dotted line on frequency histograms (A-C).

## Naturalistic sleep modulates the extent to which neural flexibility is related to visual working memory performance.

To understand the extent to which naturalistic sleep patterns moderated the relationship between node flexibility (i.e., a brain region's ability to flexibly change its affiliation across time) and visual working memory performance, we used a data driven approach and employed a linear mixed-effects analysis on every region within the Desikan-Killiany atlas parcellation. From these analyses, we found seven brain regions that showed significant ($p < .05$, uncorrected) interactions between naturalistic sleep and regional flexibility on visual working memory performance (see Table 1). Figure 2 displays these regions that included the left cuneus (l-CN), right lateral occipital (l-LO) cortex, left lingual gyrus (l-LG), left transverse temporal (l-TT) cortex, left pars triangularis (l-PT), right paracentral gyrus (r-PC), and rostral middle frontal cortex (r-RMF). Three of these regions (l-LO, l-LG, l-CN) are critical to visual function and object perception (Grill-Spector et al., 1998). Two of these regions (l-PT, r-PR) are located in the frontal cortex, which supports the cognitive demands associated with the visual working memory task (Courtney, 1998). Lastly, we also find two regions (l-TT, r-PC) that are located in the inferior temporal cortex and motor cortex, which are critical hubs in memory formation, retrieval of visual information, and deployment of action in response to this information (Chelazzi et al., 1998; Miyashita, 1993). Of these uncorrected but significant findings, a false discovery rate (FDR) correction for multiple comparisons was employed. The left cuneus and transverse temporal regions were the only brain regions to survive this correction (FDR, $p = .05$) and maintain a significant interaction on visual working memory accuracy. We also tested the robustness of these results relative to other sleep estimates by running additional analyses using an average of total sleep time for the previous 7 days and previous 13 days, but found no sleep modulation effects that survived multiple comparisons corrections with these more distal observation windows. We found that the previous night's sleep produced stronger effects on the relationship between neural flexibility and working memory, and impacted the highest number of task-relevant brain regions. As another robustness check, we ran an additional analysis excluding a potential outlier in total sleep time and found that the above results remained the same.

## Table 1. Multilevel regression model parameters for naturalistic sleep and neural flexibility on working memory performance

| | *Working Memory Performance* | | |
|---|---|---|---|
| ***predictors*** | ***coefficient*** | ***CI*** | ***p*** |
| Intercept | -0.111 | -0.601–.377 | 0.653 |
| L Cuneus | 1.445 | 0.648 – 2.24 | 0.0004** |
| Total Sleep Time (TST) | 0.001 | 0.0007 – .002 | 0.001* |
| L Cuneus X TST | -0.002 | -0.004 – -0.001 | 0.001* |
| | ***random effects*** | ***variance*** | |
| | subjects | 0.008 | |
| | sessions | 0.001 | |

| | *Working Memory Performance* | | |
|---|---|---|---|
| ***predictors*** | ***coefficient*** | ***CI*** | ***p*** |
| Intercept | 0.287 | -0.066 – 0.640 | 0.108 |
| R Paracentral Gyrus | 0.825 | 0.230 – 1.419 | .006** |
| Total Sleep Time (TST) | 0.0009 | 0.0001 – 0.001 | .021* |
| R Paracentral Gyrus X TST | -0.001 | -0.002 – 0.0003 | .018* |
| | ***random effects*** | ***variance*** | |
| | subjects | 0.009 | |
| | sessions | 0.001 | |

| | *Working Memory Performance* | | |
|---|---|---|---|
| ***predictors*** | ***coefficient*** | ***CI*** | ***p*** |
| Intercept | 0.287 | -0.262 – 0.468 | 0.111 |
| L Lingual Cortex | 0.825 | 0.552 – 1.784 | 0.007** |
| Total Sleep Time (TST) | 0.0009 | 0.0004 – 0.0002 | 0.023* |
| L Lingual X TST | -0.0016 | -0.003 – -0.0007 | 0.02* |
| | ***random effects*** | ***variance*** | |
| | subjects | 0.009 | |
| | sessions | 0.001 | |

| | *Working Memory Performance* | | |
|---|---|---|---|
| ***predictors*** | ***coefficient*** | ***CI*** | ***p*** |
| Intercept | 0.311 | -0.092 – 0.715 | 0.128 |
| L Lateral Occipital Cortex | 0.744 | .100 – 1.52 | 0.024* |
| Total Sleep Time (TST) | 0.0009 | -.0.00001 –.001 | 0.04* |
| L Lateral Occipital Cortex X TST | -0.001 | -0.003 – -.00001 | 0.037 |
| | ***random effects*** | ***variance*** | |
| | subjects | 0.009 | |
| | sessions | 0.001 | |

| | *Working Memory Performance* | | |
|---|---|---|---|
| ***predictors*** | ***coefficient*** | ***CI*** | ***p*** |
| Intercept | 0.236 | -0.170 – 0.643 | 0.251 |
| Pars Triangularis | 0.841 | 0.202 – 1.479 | 0.01* |
| Total Sleep Time (TST) | 0.001 | 0.0002 – 0.002 | 0.019* |
| Pars Triangularis X TST | -0.001 | -0.003 – -0.0003 | 0.018* |
| | ***random effects*** | ***variance*** | |
| | subjects | 0.008 | |
| | sessions | 0.001 | |

| | *Working Memory Performance* | | |
|---|---|---|---|
| ***predictors*** | ***coefficient*** | ***CI*** | ***p*** |
| Intercept | -0.355 | -0.806 – 0.094 | 0.121 |
| L Transverse Temporal Cortex | 1.763 | 1.063 – 2.462 | < .001 *** |
| Total Sleep Time (TST) | 0.002 | 0.001 – 0.003 | < .001 *** |
| L Transverse Temporal Cortex X TST | -0.003 | -0.005 – -0.002 | < .001 *** |
| | ***random effects*** | ***variance*** | |
| | subjects | 0.008 | |
| | sessions | 0.001 | |

| | *Working Memory Performance* | | |
|---|---|---|---|
| ***predictors*** | ***coefficient*** | ***CI*** | ***p*** |
| Intercept | 1.095 | 0.729 – 1.461 | < .001 *** |
| Rostral Middle Frontal Cortex | -0.55 | -1.145 – 0.044 | 0.068 |
| Total Sleep Time (TST) | -0.0009 | -0.001 – -0.0001 | 0.031* |
| Rostral Middle Frontal Cortex X TST | 0.001 | 0.0001 – 0.002 | 0.030* |
| | ***random effects*** | ***variance*** | |
| | subjects | 0.009 | |
| | sessions | 0.002 | |

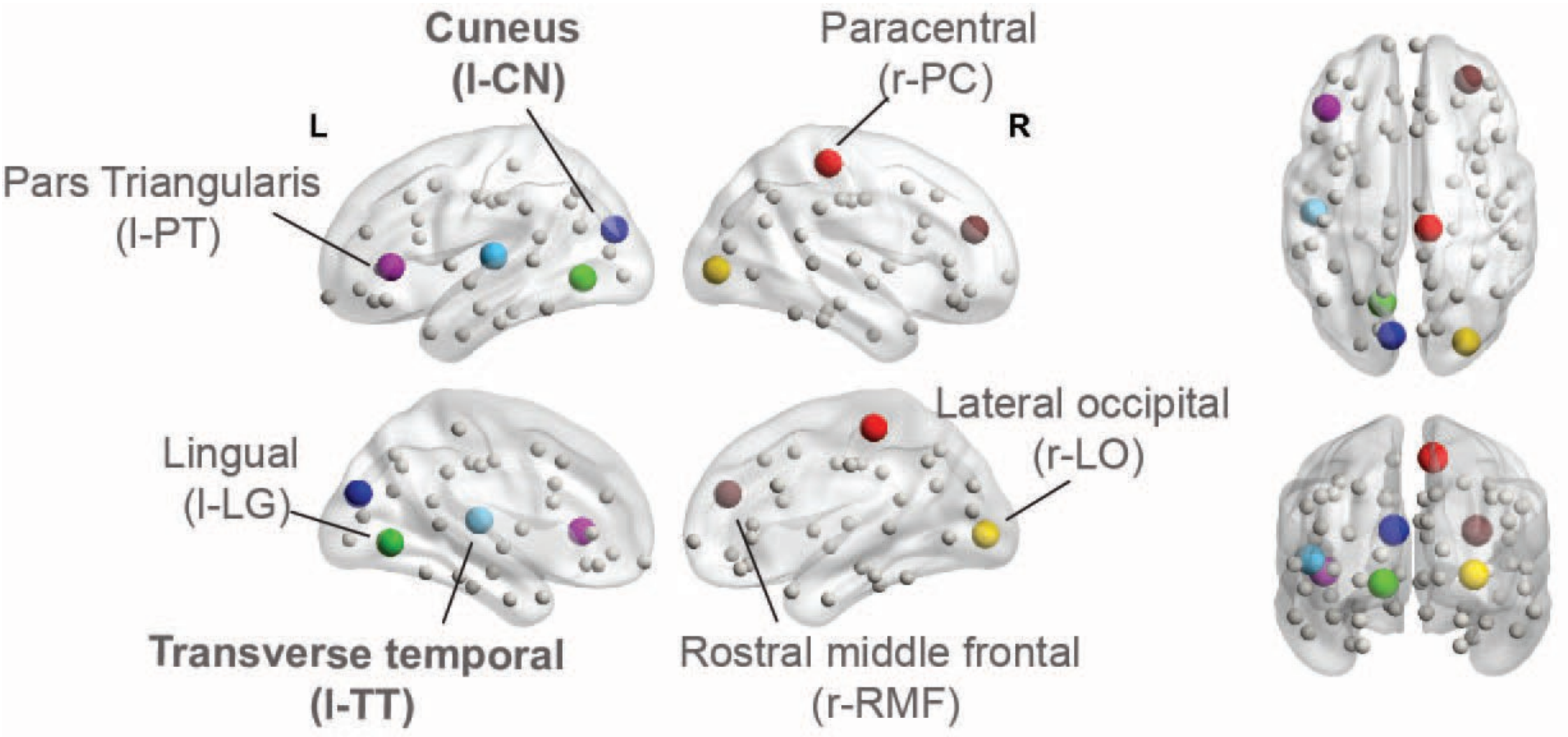

**Figure 2: Interactions between TST and regional flexibility on visual working memory performance.** Medial (top row, left), lateral (bottom row, left), ventral (top, right), and posterior (bottom, right) views of all of the brain nodes, parcellated by the Desikan-Killiany atlas (Desikan et al., 2006). Larger, non-gray nodes showed a significant (uncorrected) interaction between TST and flexibility on working memory performance. An FDR correction for multiple comparisons reveals two regions (bolded text) survived this significance thresholding (q = 0.05), the left cuneus and left transverse temporal regions.

To determine the direction of these naturalistic sleep by neural flexibility interaction effects on visual working memory performance, we conducted follow-up linear mixed effects analyses within hour-based quartiles (Arora et al., 2011; Choi et al., 2017; Patel et al., 2004; Steptoe, 2006) surrounding the mean TST (M = 451.63): less than 6 hours of sleep (N = 97), 6-7 hours of sleep (N = 126), 7-8 hours of sleep (N = 96), and greater than 8 hours of sleep (N = 121). As shown in Figure 3, we find decreasing linear trends of neural flexibility on visual working memory performance as the number of sleep hours increases, such that greatest neural flexibility effects on visual working memory accuracy are observed in the lowest sleep category, less than 6 hours of sleep.

The strongest linear trends are observed in occipital and frontal regions, specifically the left lingual gyrus, right lateral occipital cortex, left pars triangularis, and left cuneus (Figure 3A). Across all regions, we also found that the effects of neural flexibility were higher (as indicated by higher t-values) in the “less than 6 hours of sleep” category relative to the “greater than 8 hours of sleep” category (Figure 3A). Specifically, we find that the direction of the relation between neural flexibility and visual working memory performance across all levels of sleep category were positive, indicating that higher flexibility was related to higher levels of accuracy in the task (Figure 3B). When participants had the lowest levels of sleep (less than 6 hours), we found the highest number of significant positive relationships between neural flexibility and visual working memory accuracy. Importantly, in addition to this effect, across the four sleep quartiles, we did not observe any difference in behavior (Figure 3C), suggesting that even at low levels of sleep, the healthy participants in this sample were able to utilize compensatory resources. Specifically, greater flexibility in neural regions such as left cuneus, left lingual gyrus, and left pars triangularis were associated with *maintaining* higher levels of performance in the task (Figure 3B).

In support of the idea that less sleep requires more compensatory neural flexibility, we see that the next highest number of significant relationships is observed within the 6-7 hours of sleep category. Within the 6-7 hours sleep category, higher flexibility in the right paracentral gyrus, left lingual gyrus, and the right lateral occipital cortex were related to higher visual working memory accuracy. Within the 7-8 hours sleep category, we found that higher flexibility in the right paracentral gyrus was related to visual working memory accuracy (Figure 3B). Finally, we found no significant relations between neural flexibility and visual working memory performance when participants had greater than 8 hours of sleep the day prior to the testing session.

The pattern of results in Figure 3A and 3B suggests that neural flexibility is less strongly related to performance as the number of sleep hours increases, which potentially indicates that neural flexibility may serve as a “compensatory mechanism” for preserving performance under lower levels of sleep (Chee & Choo, 2004). Further support for this interpretation is found by examining visual working memory accuracy within each sleep category (Figure 3C) where we see that performance is relatively stable across all sleep categories, despite these differences in neural flexibility.

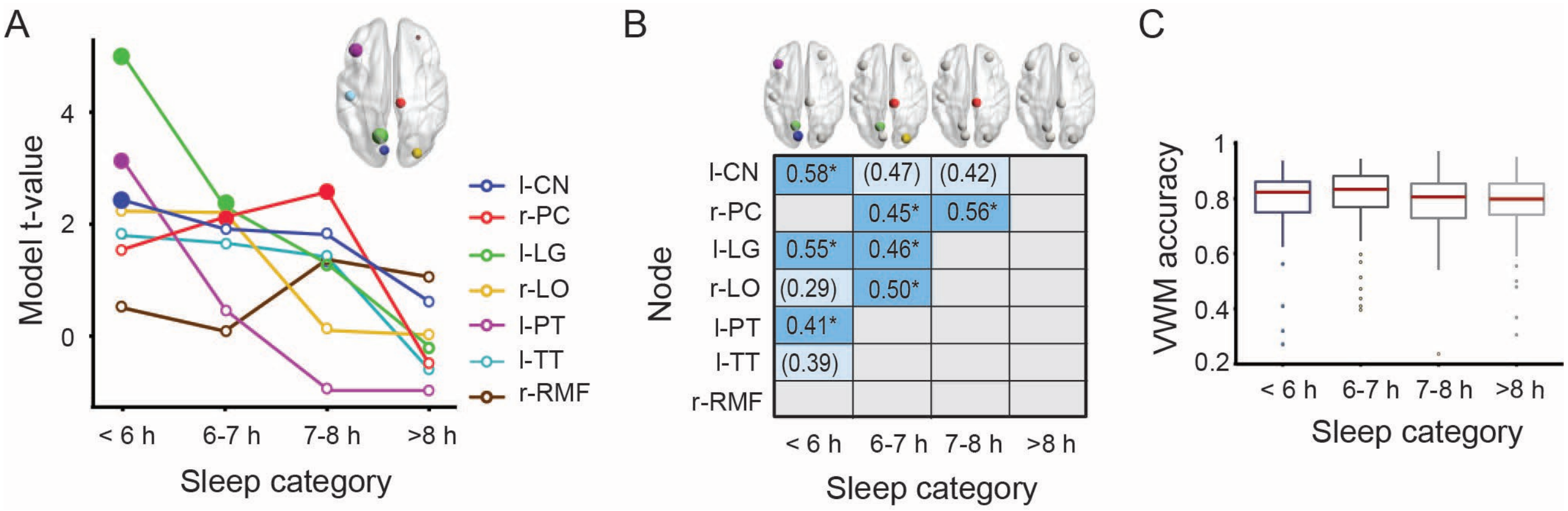

**Figure 3: Flexibility effects on visual working memory accuracy sleep quartiles.** Mixed modelling effects completed on 4 sleep categories, which encompassed low (< 6 hr) to high (>8 h) levels of sleep. (A) To probe significant interactions found in mixed effects model analysis, this line plot shows the t-value for 7 regions in which neural flexibility had a significant effect on visual working memory performance within each sleep category. Filled markers indicate significant results and brain inset displays the 7 regions with orbs plotted at the centroid scaled by the difference in t-values between < 6 h and > 8 h sleep categories. (B) Matrix of significant results in each category, sorted by the 'strength' of the compensatory effect, determined by the number of significant (blue, $p < .05$) or marginally significant (light blue, $p < .1$) effects of each region. (C) Boxplots of visual working memory performance (accuracy) across the sleep quartiles. Boxplots indicate median and 25th and 75th percentiles of the distribution. *Note:* l-CN=left cuneus, r-PC=right paracentral gyrus, l-LG=left lingual gyrus, r-LO=right lateral occipital cortex, l-PT=left pars triangularis, l-TT=left transverse temporal cortex, and r-RMF=rostral medial frontal cortex.

## Allegiance indicates which brain regions are driving the flexible nodes.

While brain network (and node) flexibility has been successfully used to characterize large-scale functional differences (e.g., Telesford et al., 2016) across a variety of tasks (Braun et al., 2015; Betzel et al., 2017), it does not, alone, afford the opportunity to inspect more granular changes in community reconfigurations. As has been previously used (Garcia et al., 2020), *allegiance*, or the proportion of time each node-pair is in the same community, may be used to describe observed network dynamics on a finer spatial scale. The granular exploration of the community changes across time may display patterns critical to the presumed compensatory increased flexibility as a result of naturalistic reductions in total sleep time.

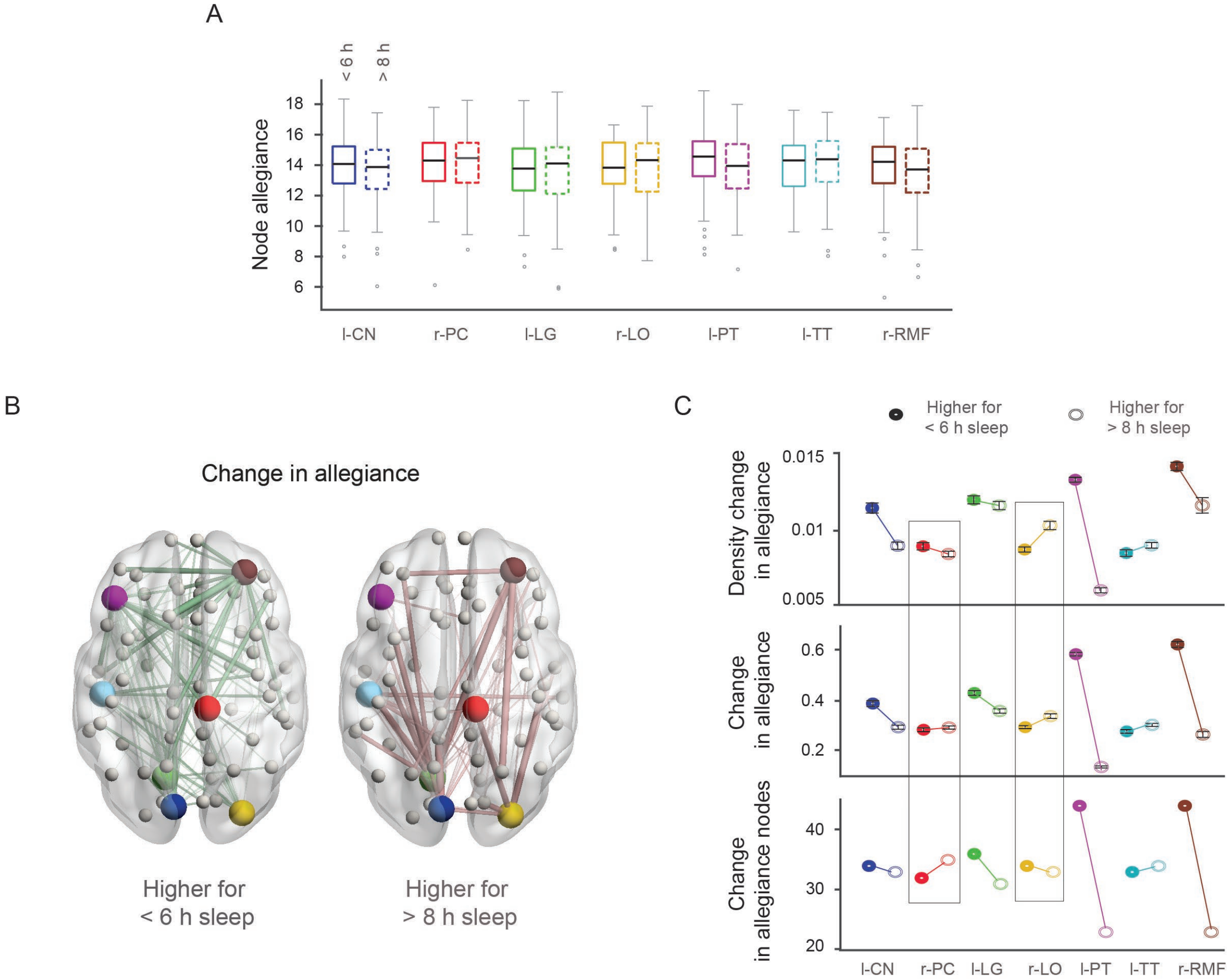

**Figure 4: Allegiances connected to each of the significant nodes**. (A) Average sum of allegiance of all nodes relative to the nodes of interest (x-axis label). Boxplots display the median and edges represent the 25th and 75th percentiles of the distribution of summed allegiance values across nodes connected to each node of interest in two sleep categories, marked by a solid (< 6h) or dotted (> 8h) box outline. (B) Spatial organization of allegiances for two sleep categories. Dorsal view of the brain with significant nodes colored. Lines connecting each node show the increases and decreases of node-pair allegiances, specifically comparing less than 6h sleep (left) to greater than 8h sleep (right). (C) Top: Density changes in allegiance, i.e., mean allegiance normalized by the number of nodes that show an increase (or decrease) in allegiance for < 6h and > 8h sleep. Error bars represent 1 standard deviation above and below the mean. Middle: Mean of allegiances that are higher for < 6h or > 8h sleep categories. Bottom: Number of nodes that on average are higher for < 6h or > 8h sleep categories. Change in density is driven by both number of nodes and average allegiances, except for two brain regions highlighted using rectangular boxes.

We specifically sought to understand whether the proposed compensatory effect of flexible nodes indicates that new networks are engaged at lower sleep levels or simply the same networks are engaged more frequently to compensate for the increased lack of sleep. While Figure 3 showed the often monotonic decreases of neural flexibility with increasing sleep for each brain region, in this section we focus on the extremes within sleep categories and concentrate on network differences between a deprived state of sleep (< 6 hours) to an abundance of sleep (> 8 hours). First, we inspected the average sum of allegiance across all sessions and subjects within these two categories (Figure 4A), and found that on average, allegiances of significant nodes do not differ significantly between the two sleep categories ($p > 0.05$ for t-test comparison). Since these nodes showed an increase in flexibility at lower sleep levels, the similarity in allegiances could imply that the node may be engaging with different nodes under different sleep conditions as defined by our sleep categories, suggesting a reconfiguration in the network structure. To confirm this, we next inspected the spatial distribution of allegiances from each of the significant nodes. In Figure 4B, we show allegiance differences between the low sleep and high sleep categories. From visual inspection, on average there are higher allegiances with occipital cortex in the higher sleep category and higher frontal allegiances in the low sleep category. In the low sleep category, we also observed a larger number of lower allegiance values, whereas allegiance values are relatively higher but fewer in number for the high sleep category. To quantify these differences, we introduce *network density*, or the average allegiance normalized by the number of nodes that shows an *increase* in the sleep category of interest. For example, in the low sleep category, we define *density* as the average allegiance between the ROI and the nodes that show a higher allegiance in the low sleep category divided by the number of nodes that are on average higher in the low sleep category in comparison to the high sleep category. Figure 4C (top panel) shows the density for each of the ROIs and we observe increases in density for each ROI for the low sleep category compared to the high sleep category except for r-LO and l-TT. The largest increases in density were observed in the l-PT and r-RMF. To confirm whether these changes in density are due to a change in the difference in the number of connections or the average allegiance, we also inspect these values separately (Figure 4C, bottom panels). In almost all ROIs, the density changes appear to be driven partially by both, where the expansion (or contraction) of the network coincides with an increase (or decrease) in allegiance. Two notable exceptions are r-PC and r-LO which show a contraction and expansion of their respective networks. Taken together with our previous findings, these results appear to indicate that at lower sleep levels, these regions are recruiting a distributed set of regions within the brain *and* making stronger regional connections to compensate for the lack of sleep.

# Discussion

Previous research has established a clear link between sleep loss and failures in memory (Rasch & Born, 2013). While much of the previous research examining memory-related failures due to poor sleep have used an acute sleep deprivation paradigm (Chee & Choo, 2004), we expanded on this work by examining unfettered sleep patterns similar to the naturalistic fluctuations in sleep that people may experience on a regular basis. We have further expanded our knowledge of the neural underpinnings of working memory showing that the flexibility of several regions in an occipital-temporal-frontal network, which may have a protective effect to sleep-related decrements in behavior. Using a mixed modelling approach (Aarts et al., 2014) coupled with large-scale brain network analytical tools, we have shown that the flexibility of regions in visual, frontal, and temporal cortex, as it is related to behavioral performance, is modulated by naturalistic fluctuations in total sleep time.

## Flexibility of brain regions compensates for poorer performance

Using a quantification of *flexibility*, we estimated how much each brain region may flexibly recombine within other modules in the brain to perform a working memory task (Bassett et al., 2011; Betzel et al., 2017; Mattar et al., 2016). While there is a long history of inspecting how regions, subregions, or systems of the brain interact to give rise to cognition (Tognoli & Kelso, 2014; Bressler & Menon, 2010; Gollo et al., 2017) - i.e., so-called functional (or effective) connectivity (Friston, 2011) - it was only recently that researchers have shown the importance of changes within these patterns of connectivity across a variety of time scales (milliseconds: Garcia et al., 2020; seconds: Shine et al., 2016; minutes: Betzel et al., 2017) are behaviorally relevant and characterize certain patient populations (Braun et al., 2016; Cooper et al., 2019).

Our results corroborate previous findings that show flexibility in the brain is correlated with executive function during a working memory task (Braun et al., 2015) and is modulated by subjective measures of fatigue and emotional state (Betzel et al., 2017). These results substantially extend prior research suggesting that flexibility in the brain may also prevent behavioral decrements associated with a potentially detrimental state (i.e., reduced sleep duration). Using a multilevel modeling approach to target the associations between flexibility in the brain and performance fluctuations across individuals, we show that the flexibility of a subset of visual, parietal, temporal, and frontal regions as they are related to working memory performance is modulated by the previous night’s sleep. Critically, however, performance on the working memory task does not change as a function of sleep. These substantial dynamic cortical changes without associated behavior change leads us to posit that the increases in flexibility may indicate a compensatory mechanism for poor sleep.

Compensatory mechanisms, or the additional recruitment of neural resources to accomplish the same task, has been observed in the literature on the aging brain (Grady, 2008), and in those with psychopathologies (Cabeza et al., 2002; Cirstea & Levin, 2000; Feigin et al., 2006; Grady et al., 2003). For example, in age related studies, increases in neural *activation* patterns have been seen in episodic memory tasks (Cabeza 2002), working memory tasks (Grady et al. 1998; Reuter-Lorenz et al. 2000), and perceptual tasks (Grady et al. 1994). Researchers have interpreted this hyperactivation in memory tasks as a compensation for age-related memory decline. In the sleep deprivation literature, links have been made between the increases in frontal lobe activity during working memory tasks in aging adults to sleep deprived young adults (Chee & Choo, 2004; Harrison & Horne, 2000), suggesting a type of compensatory neural activity even in young adults. These activity patterns, however, do not shed light upon the nature of this compensation. Could the increased activity in the eldery be a consequence of different neural populations 'working harder' to perform the same task or is another network intervening to compensate for the network failures? While our results cannot speak to age-related memory changes, our results suggest that day-to-day compensations in sleep fluctuations might be due to a combination of resource deployment and and a recruitment of other areas to maintain adequate performance in this working memory task.

## The visual cortex is critical to the compensation in working memory

Working memory is critical for many high level behaviors (Engle et al., 1999), and requires the coordination of a variety of regions including visual, parietal and frontal cortex (Gazzaley & Nobre, 2012). Our results show that the relationship between the flexibility of a distributed network of nodes and working memory performance is modulated by the previous night's sleep. This relationship was most robust in the left cuneus (l-CN), which displayed the most robust sleep-modulated effect (Figure 2, 3B), where it survived corrections for multiple comparisons and showed the largest amplitude differences between the low sleep category and most abundant. This region is a large portion of the primary visual cortex and is functionally related to basic visual processing (Grill-Spector & Malach 2004). It is also a substantial element of the *dorsal stream*, a visual pathway implicated in the representation of object locations (Goodale & Milner 1992; Laycock et al. 2011), critical to the working memory task at hand.

Our finding that the early visual cortex flexibility-performance relationship is modulated by sleep, coupled with the observation that average performance on the visual working memory task does not change as a function of sleep, suggests that the flexibility of early visual cortex may provide a protective barrier to downstream behavioral consequences of poor sleep in adults. Research investigating the neural correlates of working memory have shown that visual and parietal cortical responses are scaled by the complexity of the visual stimulus (Xu & Chun, 2006) when held within working memory and may even track the precision of the of the visual object in memory (Ester et al., 2013; Sprague et al., 2014). Our results not only show that an increase in flexibility of visual cortex is associated with a lack of sleep, but they also suggest that on average this increase in flexibility is mostly driven by more frequent affiliations with regions within it's network rather than an expansion of it's network. Our findings may indicate that if a brain's state is compromised such as due to lack of sleep, then the visual cortex may be recruited to persistently reconstruct the visual representation of a stimulus in working memory.

## Naturalistic fluctuations in sleep affect brain-behavior relationships

A unique feature of our study is that despite no manipulation of sleep, we find robust associations linking fluctuations in sleep to the associations between flexible nodes in the brain and behavior in a working memory task. While our study used a classic probe of working memory function, our experimentation measured sleep without depriving participants of sleep. This research speaks to the growing need of experimentation to leave the bench and replicate the effects “in the wild” (Niell & Stryker, 2010; Zaki & Ochsner, 2009), where we may at least partially replicate previous results in more realistic contexts that may be a better representation of our everyday experiences (Matusz et al., 2018). Our results not only replicate previous findings linking visual working memory performance to sleep quantity (Chee & Choo, 2004; Drummond et al., 2012), but also extend this literature to unfettered experimentations linking fluctuations in the previous night’s sleep to neural processes subserving key executive functions. It is important to note that we did not find substantial effects in cumulative sleep effects on neural flexibility and working memory performance when we conducted our analyses using either an average of the previous seven or thirteen days of sleep. The lack of moderator effects on brain dynamics and working memory performance using cumulative sleep measures points to the potentially large impact that a previous night’s sleep has on brain-behavior relationships.

## Potential implications of state changes in neural plasticity

It is well established that short term sleep loss has robust effects on cognition (Harrison & Horne, 2000); it is also well known that long-term sleep disturbances are associated with a variety of health conditions, mental disorders, and disease (Krystal et al., 2008). This study pairs two innovative approaches, network neuroscience and hierarchical statistical modelling, with an intensive longitudinal design over four months, to capture naturalistic fluctuations in sleep, uncovering neural changes due to a mere 25% reduction in sleep (8 hours vs 6 hours). While previous research has shown a clear but relatively modest effect of sleep deprivation on working memory performance (Chee & Choo, 2004; Drummond et al., 2012), our results show that even without severe restriction in sleep, there are still measurable changes in brain dynamics that affect how people perform the following day. This research not only expands our understanding of functional consequences of slight reductions in sleep, but it may also provide an analytical framework to probe neuroplastic changes associated with state changes. In particular, our results highlight the effects of transient sleep loss on cognition within relatively short time periods. This methodological scheme has the potential to provide insights into other potentially potent state changes in people’s everyday dynamic environments.

## Conclusions

After a poor night's sleep, the brain undergoes reconfiguration to dynamically compensate for physiological state changes to maintain performance in a challenging working memory task. This reconfiguration is characterized by both more frequent recruitment of brain regions serving normal working memory functions and an increase in the recruitment of brain regions not typically implicated in working memory functions. These results were found using a naturalistic measure of sleep, without experimental manipulation or deprivation, strengthening our understanding of sleep effects on brain dynamics and memory. We provide evidence showing a clear connection between day-to-day fluctuations in sleep prompting a reevaluation of what we mean by "normal sleep." In sum, our findings implicate neural flexibility as a potential mechanism for neural changes and negative consequences of chronic sleep loss.

## Methods

**Participants.** Forty-two participants between the ages of 18 and 35 years old (M=22.21; SD=2.99; 58% female) were recruited from the greater Santa Barbara area as part of the Cognitive Resilience and Sleep History (CRASH) research study. As part of this study, participants completed bi-weekly experimental sessions over the course of 16 weeks (see Figure 5A). Participants completed a set of five experimental tasks while functional magnetic resonance imaging (fMRI), peripheral physiology, and eye-tracking data were recorded. Additionally, daily sleep measurements of actigraphy and sleep history were collected across the 16 week study period. The present study focuses on fMRI data collected while participants completed a visual working memory task and the subjects' actigraphy data. All methods and procedures in the present study were approved by both of the Institutional Review Boards at the University of California, Santa Barbara and the U.S. Combat Capabilities Development Command Army Research Laboratory and carried out in accordance with this approved research protocol. All participants in this study provided written informed consent to participate in this approved research protocol.

**Visual working memory task.** In order to examine visual working memory performance, participants completed a standard visual working memory task (Figure 5B; Luck & Vogel, 1997) while their blood-oxygen-level-dependent (BOLD) responses were recorded using fMRI. Participants were presented either with one or six squares of different colors for 150 milliseconds, which represented the difficult and easy condition, respectively. Afterwards, a delay period of 1180 milliseconds occurred during which a fixation cross was presented. Next, participants were asked to recall whether the new presentation of either one or six colored squares was the same or different relative to the previously presented stimuli. Participants completed a total of 144 trials (72 easy, 72 difficult) over the course of three blocks of 48 trials. The task took approximately 5 minutes (302 volumes) to complete. Visual working memory performance was assessed by calculating the accuracy of total correct responses divided by the total number of trials. Accuracy was also calculated for each condition (easy, difficult) and across both conditions. On average, participants identified 78% of targets (SD=0.10) in the visual working memory task accurately (Fig 1B; Fig 5C). Paired t-test results indicated that there was a significant difference in average accuracy between the easy and difficult conditions after averaging across sessions for each subject ($t(41)=11.97$, $p < .001$). Within the easy condition, participants demonstrated a relatively high level of accuracy (M=90%, SD=0.11), whereas participants performed at a significantly lower level of accuracy within the difficult condition (M=66%; SD=0.12).

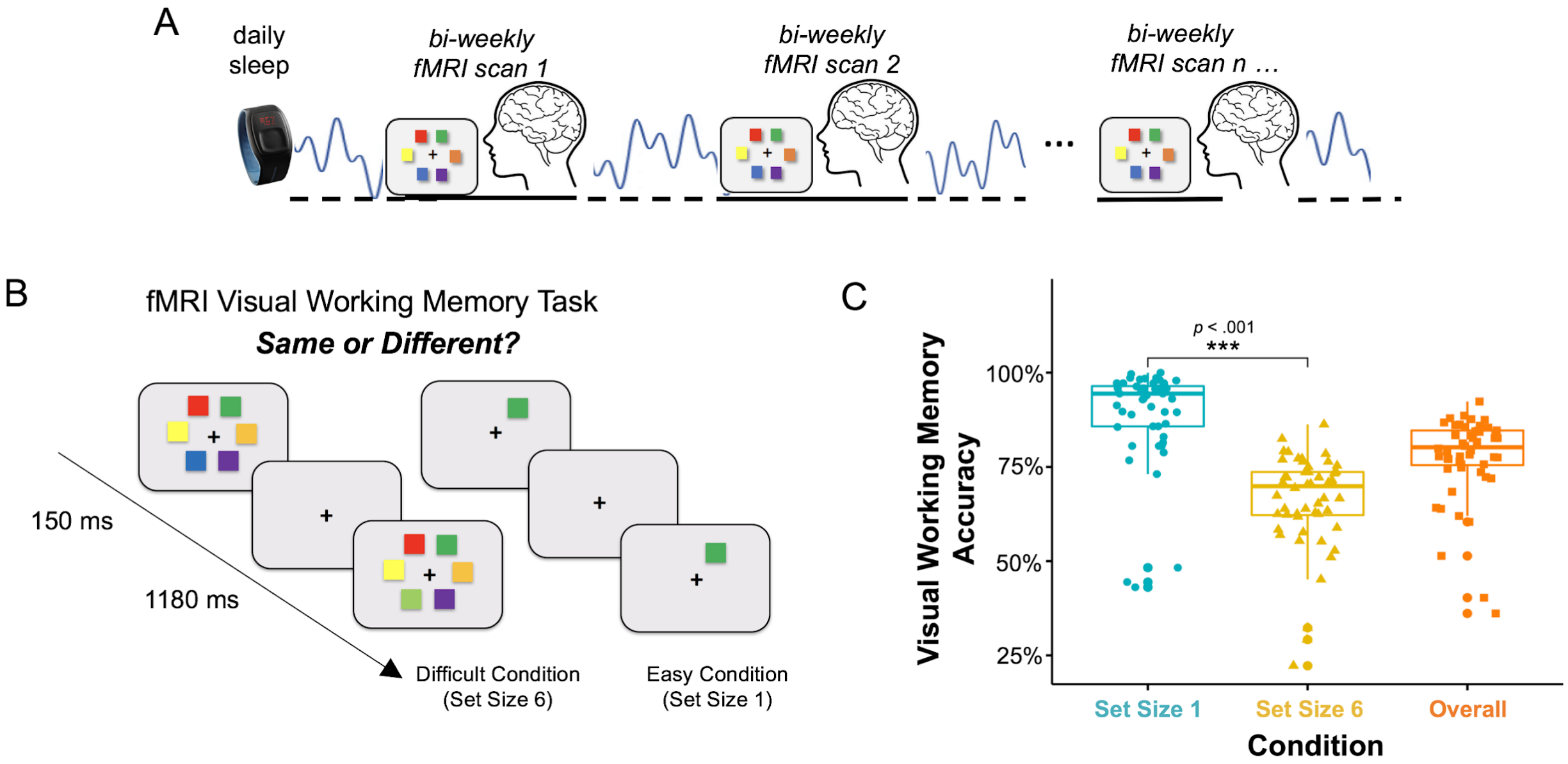

**Figure 5: Experimental paradigm.** (A) Actigraphy was used to assess sleep prior to the fMRI session. This was continuously recorded for 2-16 weeks, depending on the number of sessions for each participant. (B) The visual working memory task, derived from a traditional working memory experiment, presented subjects with an array of visual objects, and subjects were asked to determine whether a subsequently presented probe was the same or different. 1 or 6 probes were presented. (C) Performance in the working memory task, displays a difference between easy (Set Size 1) and difficult (Set Size 6) conditions.

**Naturalistic Sleep (Wrist Actigraphy & Daily Sleep Logs).** To measure fluctuations in naturalistic sleep, participants wore a Readiband Actigraph SBV2 watch (Fatigue Science, Vancouver BC) on the wrist across the 16 week study period, with the exception of removing the device during bi-weekly laboratory visits (approximately 3 hours). The actigraph device measured movement using a 3D accelerometer with a sampling rate of 16 Hz, and has been validated with respect to polysomnography and for internal consistency (Driller et al., 2016; Sadeh, 2011). Since actigraphy data were stored locally on the device, data were downloaded during the bi-weekly study sessions. The device was also charged and inspected to ensure optimal functionality. Actigraph data was preprocessed by Fatigue Science software to estimate two discrete variables at each minute for a given 24-hour period: i) individual was "in bed" or "out of bed"; and ii) individual was "asleep" or "awake." In line with previous studies (Thurman et al., 2018; Berger et al., 2008), sleep onset was defined as the first recorded instance of sleep occurring at or after 9:00pm and sleep offset was defined as the last instance of transitioning from sleep to wake before 11:00am the following day.

To obtain reliable and accurate estimates of naturalistic sleep, participants also completed a daily sleep history questionnaire. Participants completed the wake-time component of the Pittsburgh Sleep Diary (Monk et al., 1994) upon awakening each day of the study period. Specifically, participants reported the time when initiating sleep, length of time to fall asleep, awake time, number of times awake during the night, and number of minutes spent awake during awakenings. These variables allowed for subjective measurements of sleep onset, sleep offset, and wake after sleep onset, and have shown high consistency between these measurements within this particular dataset (Thurman et al., 2018). For the purpose of the present study, naturalistic sleep on the day prior to each bi-weekly session was examined to assess whether naturalistic sleep was related to either brain network dynamics or visual working memory performance. Additional analyses were conducted to examine cumulative effects of sleep by averaging naturalistic sleep before each scan session across either the previous seven or thirteen days. Using the measurements of actigraphy, total sleep time (TST) was calculated as the length of the sleep period minus the amount of the time awake during the sleep period (TST= sleep offset – sleep onset – wake after sleep onset).

**fMRI Data Acquisition.** We collected functional magnetic resonance imaging (fMRI) data using a 3T Siemens Prisma MRI. Functional image data were acquired using echo-planar imaging that involved collection of 64 coronal slices with a 3mm slice thickness, a field of view of 192 x 192 mm, flip angle of 52 degrees, and a repetition time of 910 ms. The echo time was 32 ms and the resulting voxel size of 3 x 3 x 3 mm. High-resolution structural images were also collected for coregistration and normalization of functional brain images using a magnetization prepared rapid acquisition gradient echo (MPRAGE) sequence with a spatial resolution of .9 x .9 x .9 mm and field of view of 241 x 241 mm. These structural images were collected with a repetition time of 2500 ms and an echo time of 2.22 ms.

**fMRI Preprocessing.** Neuroimaging data was preprocessed using ANTs (Avants, B. B., Tustison, N., & Song, 2009; Avants et al., 2014). The functional data underwent minimal preprocessing to correct for physiological artifacts and head motion. Physiological artifacts including respiration and cardiac cycle effects were corrected using

RETROICOR (Glover, Li, & Ress, 2000) implemented in MEAP v1.5 (Cieslak et al. 2018). Head motion was estimated using antsMotionCorr. An unbiased BOLD template was created within each session using the means of the motion-corrected BOLD time series from each run. The BOLD templates were coregistered to the corresponding T1-weighted high resolution structural images. Each session was spatially normalized to a custom study-specific multi-modal template which included T1-weighted, T2-weighted and GFA images from twenty-four randomly selected participants stratified to match the study population on gender. The template was affine transformed to share the coordinate space of the MNI152 Asymmetric template. The final BOLD time series images were created using the composed transforms from head motion correction, BOLD template coregistration, BOLD-to-T1w coregistration and spatial normalization into 3mm MNI space using a single Hamming weighted sinc interpolation. All co-registration and normalization steps were computed using ANTs.

**Functional Connectivity Analysis.** In order to examine functional connectivity in brain areas involved in visual working memory, we applied the Desikan-Killiany (DK) anatomical atlas parcellation (Desikan et al., 2006) to each subject's brain data, which divided the human cortex into 68 regions of interest (ROIs; 34 bilateral cortical regions for a total of 68 regions). To assess functional connectivity among ROIs, mean regional time-courses were extracted and standardized using the nilearn package (Abraham et al., 2014) in Python 2.7, and confound regression was then conducted. In particular, the time series for each region was detrended by regressing the time series on the mean as well as both linear and quadratic trends. There were a total of 16 confound regressors, which included: head motion, global signal, white matter, cerebrospinal fluid and derivatives, quadratics and squared derivatives. This functional connectivity preprocessing pipeline was selected based on conclusions from prior work that examined performance across multiple commonly used preprocessing pipelines for mitigating motion artifacts in functional BOLD connectivity analyses (Ciric et al., 2017; Lydon-Staley et al., 2019). Following confound regression, wavelet coherence was estimated for each pair of regions, and was averaged across frequency bands between .06 and .12 Hz, a task-relevant frequency range of coherence (Sun, Miller, & D'Esposito, 2004) within 20 sec windows, yielding a 68 x 68 matrix of coherence values for each pair of regions for each time window.

**Community Detection and Network Dynamics Metrics (flexibility and allegiance).** While human brain mapping efforts have demonstrated a relationship between spatial specificity and cognitive functions, techniques rooted in network science provide a useful framework for characterizing and understanding the spatiotemporal dynamics of the functional systems subserving cognition (Bassett & Sporns, 2017). One of the core concepts at the basis of network science is network modularity, which is the idea that neural units are structurally or functionally connected forming modules or clusters (Garcia et al.,2018). This organization allows for the system to perform both local-level exchanges of information, while maintaining system-level performance. Here, we examine whether interactions with network communities (i.e., flexibility) for each region was related to visual working memory performance, and how these relations were modulated by naturalistic fluctuations in sleep. To measure such changes in network communities during the visual memory task, a multilayer community detection analysis was employed (Bassett et al., 2011; Mucha et al., 2010). In particular, we utilized a Louvain algorithm to maximize modularity (Blondel et al., 2008) to define functional communities, and this optimization procedure was repeated 100 times, since the algorithm is susceptible to multiple solutions (Good et al., 2010). From these multiple iterations, two community metrics were computed: (i) *flexibility*, or proportion of time during which each node switches to a different community assignment; and (ii) *allegiance*, or the proportion of time that a pair of nodes were assigned to the same community.

In more concrete terms, the flexibility of each node corresponds to the number of instances in which a node changes community affiliation, *g*, normalized by the total possible number of changes that could occur across the layers *L*. In other words, the flexibility of a single node *i*, $\xi_i$, may be estimated with

$$\xi i = \frac{g_i}{L-1}, \quad (1)$$

where L is the total number of temporal windows. Flexibility, in these terms, is a node-level metric of community dynamics across time.

Allegiance, on the other, is a nodal pair-wise estimate estimating the amount of time each pair of nodes spends in the same community. We define the allegiance matrix *P*, where edge weight $P_{ij}$ denotes the number of times a pair of nodes moves to the same community together divided by $L-1$ possible changes. Allegiance, in contrast to flexibility, is a more granular estimate of community dynamics, allowing a pair-wise estimate of community affiliations.

**Data Analysis.** To test whether the relation between flexibility within brain regions of interest and visual working memory performance were modulated by naturalistic fluctuations in sleep, mixed-effects analyses were conducted using the *lme4* package in R (Bates et al., 2015), and p-values were obtained using the *sjstats* (Ludecke, 2020) package. Mixed-effects analyses allow for accurate specification of hierarchical data structures (Bryk & Raudenbush, 1992), such as accounting for repeated measures within subjects and between-subjects session level effects, which strengthens our ability to make valid statistical inferences.

Using a data-driven approach, we constructed a separate multilevel model for each brain region of interest to test whether naturalistic sleep (total sleep time on the day prior to the testing session) moderated the relation between node flexibility and visual working memory accuracy. Specifically, fixed factors included naturalistic sleep (i.e., total sleep time), neural flexibility of a single brain region, and the interaction of naturalistic sleep and neural flexibility. We also included random intercepts of subject and session in each model to account for between-subjects variability and session-level changes across time. All predictors in each model were grand mean-centered. This resulted in a total of 68 models (See *Supplementary* materials). To determine whether the inclusion of a random effect of session produced an improved model fit to the observed data, a Likelihood Ratio Test was conducted to compare $X^2$ differences between models. Significantly larger $X^2$ differences and thus, lower Akaike Information Criterion (AIC) values indicated the best fitting model.

Each model was specified as follows:

$$OverallAccuracy_{ij} = \beta_0 + \beta_1 * NodeFlexibility_{1ij} + \beta_2 * Sleep_{2ij} + \beta_3 * NodeFlexibilityXSleep_{3ij} + u_i + v_j + e_{ij}$$

Where i= subject and j= session. The random intercepts for subject and session are represented by $u_i$ and $v_j$, respectively. Given the large number of models tested, p-values were adjusted using false discovery rate (FDR) multiple comparisons correction.

For models where a significant interaction between naturalistic sleep and neural flexibility was found, interactions were probed using a priori sleep categories: less than 6 hours of sleep (*N*=97), 6-7 hours of sleep (*N*=126), 7-8 hours of sleep (*N*=96), and greater than 8 hours of sleep (*N*=121; Arora et al., 2011; Choi et al., 2017; Patel et al., 2004; Steptoe, 2006). A linear mixed effects analysis was conducted to test the effects of neural flexibility and visual working memory accuracy for each sleep category (Fig 3A & 3B). The following factors were included in the model: fixed factor of neural flexibility, random intercept of subject, and random intercept of session. Similar to our analysis procedure in our primary analysis, we conducted a Likelihood Ratio Test to determine whether the inclusion of a random effect of session produced an improved model fit to the observed data relative to the model that included only a random intercept for each subject. Based on this analysis, the model that provided the best fit to the observed data as indicated by a significant $X^2$ difference and lower AIC value was used for interpretation.

## Acknowledgements

This research is only one small segment of a larger study funded by the US CCDC Army Research Laboratory, but carried out by collaborators at the University of California, Santa Barbara as part of the Institute for Collaborative Biotechnologies. As part of this large project, the authors would like express gratitude to those that have in any way contributed to the dataset including Nick Wasylyshyn, Steven Tompson, Matthew Cieslak, Greg Lieberman, Heather Roy, and to those that contributed by study coordination and subject testing including Phil Beach, Mario Mendoza, Hannah Erro, Gold Okafor, Alex Asturias and Zoe Rathbun. Finally, we would like to thank Mary Zhuo Ke for her assistance in reviewing the manuscript prior to submission. The views and conclusions contained in this document are those of the authors and should not be interpreted as representing the official policies, either expressed or implied, of the US CCDC Army Research Laboratory or the US Government. E.B.F. acknowledges support from ARL (Cooperative Agreement Number W911NF-10-2-0022, Subcontract Number APX02-0006, PI Bassett), DARPA (Award Numbers: 140D0419C0093, FA8650-17-C-7712) and ARO (ARO W911NF1810244), and NIH/NCI 1R01CA229305-01A1.

## Author Contributions

N.L. led analysis. N.L., J.O.G., K.B. developed framework and modeling approach. N.L., K.B., J.O.G., E.F.E., E.F., S.T. provided input on data analysis. N.L., J.O.G., K.B. conducted analyses. N.L., S.T., K.B., J.O.G. wrote the manuscript. B.G., S.G., J.C.E., J.M.V. planned the experimentation and oversaw the data collection. J.O.G, E.F. supervised the research. N.L., K.B. prepared the figures and tables. B.G., S.G., E.F., E.F.E., J.O.G., N.L., S.T., K.B. all reviewed and provided feedback on the submitted manuscript.

## Supplementary Material

# Flexibility of brain regions during working memory curtails cognitive consequences to lack of sleep

**Nina Lauharatanahirun[1,2,6,7*], Kanika Bansal[1,3], Steven M. Thurman[1], Jean M. Vettel[1,4], Barry Giesbrecht[4,8], Scott Grafton[4,8], James C. Elliott[4,8], Erin Flynn-Evans[5], Emily Falk[2] & Javier O. Garcia[1*]**

[1]Human Research and Engineering Directorate, US Combat Capabilities Development Command Army Research Laboratory, Aberdeen Proving Ground, MD
[2]Annenberg School for Communication, University of Pennsylvania, Philadelphia, PA
[3]Department of Biomedical Engineering, Columbia University, New York, NY
[4]Psychology and Brain Sciences, University of California, Santa Barbara, Santa Barbara, CA
[5]National Aeronautics and Space Administration Ames Research Center, Mountain View, CA
[6]Department of Biomedical Engineering, Pennsylvania State University, State College, PA
[7]Department of Biobehavioral Health, Pennsylvania State University, State College, PA
[8]Institute for Collaborative Biotechnologies, University of California, Santa Barbara, Santa Barbara, CA
*Correspondence to nina.lauhara@gmail.com or javiomargarcia@gmail.com

**Supplementary Table 1. Multilevel model parameters for naturalistic sleep (TST) and neural flexibility on working memory performance**

| *Brain Node Name* | *Intercept B* | *Intercept SE* | *Intercept t* | *Intercept p* | *Brain Node B* | *Brain Node SE* | *Brain Node t* | *Brain Node p* | *TST B* | *TST SE* | *TST t* | *TST p* | *TST X Brain Node B* | *TST X Brain Node SE* | *TST X Brain Node t* | *TST X Brain Node p* | *TST X Brain Node $p_{corrected}$* |
|---|---|---|---|---|---|---|---|---|---|---|---|---|---|---|---|---|---|
| **L Superior Temporal Sulcus** | 0.555 | 0.240 | 2.308 | 0.022 * | 0.337 | 0.377 | 0.893 | 0.372 | 0.0003 | 0.0005 | 0.577 | 0.564 | -0.0005 | 0.0008 | -0.597 | 0.551 | 0.936 |
| **R Superior Temporal Sulcus** | 0.879 | 0.200 | 4.393 | <.001 *** | -0.184 | 0.319 | -0.578 | 0.563 | -0.0003 | 0.0004 | -0.901 | 0.368 | 0.0006 | 0.0007 | 0.907 | 0.365 | 0.828 |
| **L Caudal Anterior Cingulate Cortex** | 0.463 | 0.209 | 2.208 | 0.028 * | 0.477 | 0.329 | 1.451 | 0.148 | 0.0004 | 0.0004 | 0.911 | 0.363 | -0.0006 | 0.0007 | -0.911 | 0.363 | 0.828 |
| **R Caudal Anterior Cingulate Cortex** | 0.295 | 0.198 | 1.483 | 0.139 | 0.768 | 0.320 | 2.396 | 0.017 * | 0.0006 | 0.0004 | 1.427 | 0.155 | -0.001 | 0.0007 | -1.449 | 0.148 | 0.632 |
| **L Caudal Middle Frontal Cortex** | 0.720 | 0.214 | 3.362 | <.001 *** | 0.074 | 0.341 | 0.218 | 0.827 | 0.0001 | 0.0004 | 0.268 | 0.788 | -0.0002 | 0.0007 | -0.280 | 0.779 | 0.978 |
| **R Caudal Middle Frontal Cortex** | 0.928 | 0.151 | 6.130 | <.001 *** | -0.276 | 0.251 | -1.097 | 0.273 | -0.0003 | 0.0003 | -1.121 | 0.263 | 0.0006 | 0.0005 | 1.123 | 0.262 | 0.707 |
| **L Cuneus** | -0.111 | 0.246 | -0.453 | 0.650 | 1.445 | 0.401 | 3.600 | <.001 *** | 0.001 | 0.0005 | 3.242 | 0.001 ** | -0.002 | 0.0008 | -3.274 | 0.001 ** | 0.042 * |
| **R Cuneus** | 0.686 | 0.186 | 3.674 | <.001 *** | 0.134 | 0.310 | 0.432 | 0.666 | -0.00001 | 0.0004 | -0.033 | 0.973 | 0.00001 | 0.0007 | 0.025 | 0.979 | 0.994 |
| **L Entorhinal Cortex** | 0.548 | 0.215 | 2.547 | 0.011 * | 0.306 | 0.301 | 1.016 | 0.310 | 0.0003 | 0.0004 | 0.683 | 0.495 | -0.0004 | 0.0006 | -0.670 | 0.503 | 0.904 |
| **R Entorhinal Cortex** | 0.624 | 0.273 | 2.282 | 0.023 * | 0.205 | 0.383 | 0.535 | 0.592 | 0.0004 | 0.0006 | 0.745 | 0.457 | -0.0006 | 0.0008 | -0.764 | 0.445 | 0.904 |
| **L Frontal Pole** | 0.512 | 0.241 | 2.120 | 0.035 * | 0.352 | 0.342 | 1.030 | 0.304 | 0.0008 | 0.0005 | 1.575 | 0.116 | -0.001 | 0.0007 | -1.574 | 0.117 | 0.573 |
| **R Frontal Pole** | 0.854 | 0.249 | 3.420 | <.001 *** | -0.130 | 0.359 | -0.363 | 0.716 | -0.00005 | 0.0005 | -0.100 | 0.920 | 0.00007 | 0.0007 | 0.096 | 0.923 | 0.994 |
| **L Fusiform Gyrus** | 0.470 | 0.231 | 2.033 | 0.043 * | 0.464 | 0.358 | 1.296 | 0.196 | 0.0006 | 0.0005 | 1.180 | 0.239 | -0.001 | 0.0008 | -1.198 | 0.232 | 0.690 |
| **R Fusiform Gyrus** | 0.912 | 0.187 | 4.864 | <.001 *** | -0.226 | 0.292 | -0.773 | 0.440 | -0.0004 | 0.0004 | -1.004 | 0.316 | 0.0006 | 0.0006 | 1.001 | 0.318 | 0.772 |
| **L Inferior Parietal Cortex** | 0.800 | 0.188 | 4.242 | <.001 *** | -0.059 | 0.316 | -0.188 | 0.850 | -0.0002 | 0.0004 | -0.650 | 0.516 | 0.0004 | 0.0006 | 0.646 | 0.518 | 0.904 |
| **R Inferior Parietal Cortex** | 0.711 | 0.172 | 4.114 | <.001 *** | 0.090 | 0.305 | 0.297 | 0.766 | -0.00007 | 0.0003 | -0.203 | 0.839 | 0.0001 | 0.0006 | 0.201 | 0.840 | 0.978 |
| **L Inferior Temporal Cortex** | 0.746 | 0.177 | 4.202 | <.001 *** | 0.030 | 0.257 | 0.118 | 0.905 | 0.00004 | 0.0003 | 0.118 | 0.906 | -0.00007 | 0.0005 | -0.131 | 0.895 | 0.994 |
| **R Inferior Temporal Cortex** | 0.843 | 0.177 | 4.757 | <.001 *** | -0.122 | 0.269 | -0.454 | 0.650 | -0.0001 | 0.0004 | -0.259 | 0.795 | 0.0001 | 0.0006 | 0.258 | 0.796 | 0.978 |
| **L Insular Cortex** | 0.621 | 0.156 | 3.965 | <.001 *** | 0.245 | 0.263 | 0.930 | 0.353 | 0.00005 | 0.0003 | 0.171 | 0.863 | -0.0001 | 0.0005 | -0.174 | 0.861 | 0.978 |
| **R Insular Cortex** | 0.723 | 0.147 | 4.908 | <.001 *** | 0.074 | 0.260 | 0.287 | 0.774 | -0.0001 | 0.0003 | -0.331 | 0.740 | 0.0001 | 0.0005 | 0.316 | 0.752 | 0.978 |
| **L Isthmus Cingulate Gyrus** | 0.861 | 0.296 | 2.905 | 0.004 ** | -0.131 | 0.425 | -0.310 | 0.756 | -0.0001 | 0.0006 | -0.196 | 0.844 | 0.0001 | 0.0009 | 0.178 | 0.858 | 0.978 |
| **R Isthmus Cingulate Gyrus** | 0.278 | 0.214 | 1.297 | 0.196 | 0.736 | 0.324 | 2.271 | 0.024 * | 0.0008 | 0.0004 | 1.740 | 0.083 | -0.001 | 0.0007 | -1.725 | 0.086 | 0.562 |
| **L Lateral Occipital Cortex** | 0.619 | 0.201 | 3.080 | 0.002 ** | 0.239 | 0.323 | 0.739 | 0.460 | 0.0001 | 0.0004 | 0.423 | 0.672 | -0.0003 | 0.0007 | -0.428 | 0.669 | 0.967 |
| **R Lateral Occipital Cortex** | 0.311 | 0.203 | 1.528 | 0.128 | 0.744 | 0.327 | 2.274 | 0.024 * | 0.0009 | 0.0004 | 2.065 | 0.040 * | -0.001 | 0.0007 | -2.092 | 0.037 * | 0.366 |
| **L Lateral Orbitofrontal Cortex** | 0.715 | 0.223 | 3.204 | 0.001 ** | 0.071 | 0.334 | 0.212 | 0.831 | 0.0002 | 0.0004 | 0.498 | 0.619 | -0.0003 | 0.0007 | -0.504 | 0.614 | 0.951 |
| **R Lateral Orbitofrontal Cortex** | 0.833 | 0.180 | 4.614 | <.001 *** | -0.103 | 0.265 | -0.389 | 0.697 | -0.00007 | 0.0004 | -0.176 | 0.860 | 0.0001 | 0.0006 | 0.172 | 0.863 | 0.978 |
| **L Lingual Gyrus** | 0.287 | 0.177 | 1.614 | 0.108 | 0.825 | 0.299 | 2.757 | 0.006 ** | 0.0009 | 0.0003 | 2.311 | 0.021 * | -0.0015 | 0.0006 | -2.372 | 0.018 * | 0.255 |
| **R Lingual Gyrus** | 0.585 | 0.151 | 3.867 | <.001 *** | 0.331 | 0.274 | 1.210 | 0.227 | 0.0002 | 0.0003 | 0.642 | 0.521 | -0.0004 | 0.0006 | -0.671 | 0.502 | 0.904 |
| **L Medial Orbitofrontal Cortex** | 0.546 | 0.170 | 3.205 | 0.001 ** | 0.335 | 0.252 | 1.328 | 0.185 | 0.0004 | 0.0004 | 1.220 | 0.223 | -0.0007 | 0.0005 | -1.251 | 0.212 | 0.690 |
| **R Medial Orbitofrontal Cortex** | 0.569 | 0.189 | 3.001 | 0.003 ** | 0.293 | 0.278 | 1.055 | 0.292 | 0.0004 | 0.0004 | 1.118 | 0.265 | -0.0007 | 0.0006 | -1.140 | 0.255 | 0.707 |
| **L Middle Temporal Cortex** | 0.715 | 0.214 | 3.334 | 0.001 ** | 0.085 | 0.345 | 0.247 | 0.804 | 0.0001 | 0.0004 | 0.263 | 0.792 | -0.0002 | 0.0007 | -0.274 | 0.783 | 0.978 |
| **R Middle Temporal Cortex** | 0.924 | 0.156 | 5.898 | <.001 *** | -0.261 | 0.250 | -1.044 | 0.297 | -0.0002 | 0.0003 | -0.636 | 0.525 | 0.0003 | 0.0005 | 0.649 | 0.516 | 0.904 |
| **L Paracentral Gyrus** | 0.791 | 0.270 | 2.922 | 0.003 ** | -0.034 | 0.401 | -0.086 | 0.931 | 0.00001 | 0.0006 | 0.027 | 0.977 | -0.00003 | 0.0008 | -0.038 | 0.969 | 0.994 |
| **R Paracentral Gyrus** | 0.054 | 0.245 | 0.220 | 0.825 | 1.101 | 0.376 | 2.926 | 0.003 ** | 0.001 | 0.0005 | 2.564 | 0.011 * | -0.002 | 0.0008 | -2.582 | 0.010 * | 0.241 |
| **L Parahippocampal Gyrus** | 0.504 | 0.263 | 1.913 | 0.057 † | 0.378 | 0.370 | 1.021 | 0.308 | 0.0003 | 0.0005 | 0.530 | 0.596 | -0.0004 | 0.0008 | -0.552 | 0.581 | 0.940 |
| **R Parahippocampal Gyrus** | 0.485 | 0.233 | 2.080 | 0.038 * | 0.401 | 0.329 | 1.217 | 0.224 | 0.0007 | 0.0005 | 1.338 | 0.182 | -0.001 | 0.0007 | -1.354 | 0.177 | 0.634 |
| **L Pars Opercularis** | 0.447 | 0.208 | 2.146 | 0.033 * | 0.497 | 0.321 | 1.547 | 0.123 | 0.0006 | 0.0004 | 1.372 | 0.171 | -0.0009 | 0.0007 | -1.379 | 0.169 | 0.634 |
| **R Pars Opercularis** | 0.719 | 0.253 | 2.839 | 0.005 ** | 0.074 | 0.402 | 0.185 | 0.852 | 0.00003 | 0.0005 | 0.059 | 0.952 | -0.00005 | 0.0008 | -0.062 | 0.950 | 0.994 |
| **L Pars Orbitalis** | 0.687 | 0.284 | 2.416 | 0.016 * | 0.119 | 0.421 | 0.283 | 0.776 | 0.0003 | 0.0006 | 0.561 | 0.574 | -0.0005 | 0.0009 | -0.571 | 0.568 | 0.940 |
| **R Pars Orbitalis** | 0.677 | 0.198 | 3.405 | <.001 *** | 0.125 | 0.292 | 0.429 | 0.667 | 0.0003 | 0.0004 | 0.744 | 0.457 | -0.0004 | 0.0006 | -0.747 | 0.455 | 0.904 |
| **L Pars Triangularis** | 0.236 | 0.205 | 1.150 | 0.251 | 0.840 | 0.322 | 2.606 | 0.010 * | 0.001 | 0.0004 | 2.368 | 0.019 * | -0.001 | 0.0007 | -2.378 | 0.018 * | 0.255 |
| **R Pars Triangularis** | 0.801 | 0.199 | 4.013 | <.001 *** | -0.059 | 0.331 | -0.180 | 0.856 | -0.0001 | 0.0004 | -0.282 | 0.778 | 0.0002 | 0.0007 | 0.275 | 0.783 | 0.978 |
| **L Pericalcarine Cortex** | 0.518 | 0.160 | 3.238 | 0.001 ** | 0.419 | 0.266 | 1.577 | 0.116 | 0.0004 | 0.0003 | 1.182 | 0.238 | -0.0007 | 0.0006 | -1.204 | 0.229 | 0.690 |
| **R Pericalcarine Cortex** | 0.465 | 0.175 | 2.656 | 0.008 ** | 0.533 | 0.304 | 1.751 | 0.081 | 0.0005 | 0.0003 | 1.527 | 0.128 | -0.001 | 0.0006 | -1.548 | 0.123 | 0.573 |
| **L Postcentral Gyrus** | 0.444 | 0.145 | 3.047 | 0.002 ** | 0.580 | 0.254 | 2.278 | 0.023 * | 0.0005 | 0.0003 | 1.684 | 0.093 | -0.0009 | 0.0005 | -1.752 | 0.081 | 0.562 |
| **R Postcentral Gyrus** | 0.460 | 0.146 | 3.149 | 0.001 ** | 0.552 | 0.257 | 2.145 | 0.033 * | 0.0005 | 0.0003 | 1.597 | 0.112 | -0.0009 | 0.0005 | -1.642 | 0.102 | 0.573 |
| **L Posterior Cingulate Cortex** | 0.367 | 0.233 | 1.579 | 0.116 | 0.612 | 0.352 | 1.736 | 0.084 | 0.0008 | 0.0005 | 1.511 | 0.132 | -0.001 | 0.0008 | -1.535 | 0.126 | 0.573 |
| **R Posterior Cingulate Cortex** | 0.467 | 0.214 | 2.183 | 0.030 * | 0.477 | 0.338 | 1.411 | 0.159 | 0.0004 | 0.0004 | 1.072 | 0.284 | -0.0007 | 0.0007 | -1.080 | 0.281 | 0.708 |
| **L Precentral Gyrus** | 0.668 | 0.142 | 4.681 | <.001 *** | 0.171 | 0.251 | 0.681 | 0.496 | 0.00005 | 0.0003 | 0.180 | 0.857 | -0.0001 | 0.0005 | -0.177 | 0.859 | 0.978 |
| **R Precentral Gyrus** | 0.661 | 0.155 | 4.256 | <.001 *** | 0.188 | 0.268 | 0.701 | 0.484 | 0.0002 | 0.0003 | 0.656 | 0.512 | -0.0003 | 0.0005 | -0.679 | 0.497 | 0.904 |
| **L Precuneus** | 0.583 | 0.191 | 3.044 | 0.002 ** | 0.297 | 0.315 | 0.943 | 0.347 | 0.0002 | 0.0004 | 0.493 | 0.622 | -0.0003 | 0.0006 | -0.482 | 0.629 | 0.951 |
| **R Precuneus** | 0.576 | 0.174 | 3.309 | 0.001 ** | 0.343 | 0.313 | 1.094 | 0.275 | 0.0001 | 0.0003 | 0.342 | 0.732 | -0.0002 | 0.0006 | -0.353 | 0.723 | 0.978 |
| **L Rostral Anterior Cingulate Cortex** | 0.754 | 0.229 | 3.286 | 0.001 ** | 0.016 | 0.336 | 0.049 | 0.960 | -0.0001 | 0.0004 | -0.219 | 0.826 | 0.0001 | 0.0006 | 0.225 | 0.822 | 0.978 |
| **R Rostral Anterior Cingulate Cortex** | 0.840 | 0.262 | 3.198 | 0.001 ** | -0.109 | 0.392 | -0.277 | 0.781 | -0.000007 | 0.0005 | -0.012 | 0.989 | -0.0000004 | 0.0008 | 0.000 | 0.999 | 0.999 |
| **L Rostral Middle Frontal Cortex** | 0.501 | 0.185 | 2.708 | 0.007 ** | 0.443 | 0.308 | 1.435 | 0.152 | 0.0003 | 0.0004 | 0.870 | 0.385 | -0.0005 | 0.0006 | -0.861 | 0.390 | 0.855 |
| **R Rostral Middle Frontal Cortex** | 1.095 | 0.184 | 5.932 | <.001 *** | -0.550 | 0.299 | -1.834 | 0.068 | -0.0009 | 0.0004 | -2.162 | 0.031 * | 0.001 | 0.0006 | 2.187 | 0.029 * | 0.339 |
| **L Superior Frontal Gyrus** | 0.514 | 0.174 | 2.945 | 0.003 ** | 0.432 | 0.291 | 1.483 | 0.139 | 0.0005 | 0.0003 | 1.329 | 0.185 | -0.0008 | 0.0006 | -1.361 | 0.174 | 0.634 |
| **R Superior Frontal Gyrus** | 0.687 | 0.150 | 4.554 | <.001 *** | 0.144 | 0.258 | 0.558 | 0.577 | 0.0001 | 0.0003 | 0.318 | 0.750 | -0.0002 | 0.0005 | -0.352 | 0.725 | 0.978 |
| **L Superior Parietal Gyrus** | 0.726 | 0.195 | 3.721 | <.001 *** | 0.066 | 0.313 | 0.213 | 0.831 | 0.00003 | 0.0004 | 0.085 | 0.931 | -0.00006 | 0.0007 | -0.092 | 0.926 | 0.994 |
| **R Superior Parietal Gyrus** | 0.924 | 0.168 | 5.476 | <.001 *** | -0.277 | 0.291 | -0.953 | 0.341 | -0.0004 | 0.0003 | -1.103 | 0.271 | 0.0007 | 0.0006 | 1.105 | 0.270 | 0.707 |
| **L Superior Temporal Gyrus** | 0.410 | 0.152 | 2.695 | 0.007 ** | 0.635 | 0.262 | 2.419 | 0.016 * | 0.0005 | 0.0003 | 1.721 | 0.087 | -0.0009 | 0.0005 | -1.786 | 0.075 | 0.562 |
| **R Superior Temporal Gyrus** | 0.498 | 0.168 | 2.961 | 0.003 ** | 0.478 | 0.294 | 1.622 | 0.106 | 0.0004 | 0.0003 | 1.171 | 0.243 | -0.0007 | 0.0006 | -1.195 | 0.233 | 0.690 |
| **L Supramarginal Gyrus** | 0.694 | 0.169 | 4.088 | <.001 *** | 0.120 | 0.294 | 0.408 | 0.683 | -0.00001 | 0.0003 | -0.039 | 0.968 | 0.00003 | 0.0006 | 0.046 | 0.963 | 0.994 |
| **R Supramarginal Gyrus** | 0.830 | 0.154 | 5.375 | <.001 *** | -0.113 | 0.270 | -0.421 | 0.673 | -0.0001 | 0.0003 | -0.487 | 0.626 | 0.0003 | 0.0006 | 0.482 | 0.629 | 0.951 |
| **L Temporal Pole** | 0.903 | 0.278 | 3.249 | 0.001 ** | -0.194 | 0.394 | -0.494 | 0.621 | -0.0002 | 0.0006 | -0.434 | 0.664 | 0.0003 | 0.0008 | 0.428 | 0.668 | 0.967 |
| **R Temporal Pole** | 0.951 | 0.237 | 4.010 | <.001 *** | -0.264 | 0.338 | -0.782 | 0.434 | -0.0004 | 0.0005 | -0.798 | 0.425 | 0.0006 | 0.0007 | 0.794 | 0.428 | 0.904 |
| **L Transverse Temporal Cortex** | -0.355 | 0.227 | -1.563 | 0.119 | 1.762 | 0.353 | 4.986 | <.001 *** | 0.002 | 0.0005 | 4.397 | <.001 *** | -0.003 | 0.0007 | -4.454 | <.001 *** | 0.001 ** |
| **R Transverse Temporal Cortex** | 0.179 | 180.337 | 0.086 | 0.099 | 0.291 | 177.329 | 0.011 | 0.131 | 0.0003 | 184.434 | 0.089 | 0.580 | 0.0006 | 184.889 | 0.091 | 0.562 | 0.308 |

*Note: * < .05; ** < .01; *** < .001*

**Supplementary Table 2. Multilevel model parameters for neural flexibility on working memory performance within 4 sleep categories**

***Less than 6 hours of sleep***

| *Brain Node* | *B* | *SE* | *t* | *p* |
|---|---|---|---|---|
| L Cuneus | 0.584 | 0.24 | 2.43 | 0.024* |
| L Lateral Occipital Cortex | 0.287 | 0.128 | 2.23 | 0.041* |
| L Lingual Gyrus | 0.551 | 0.109 | 5.015 | < .001*** |
| R Paracentral Gyrus | 0.239 | 0.154 | 1.549 | 0.14 |
| L Pars Triangularis | 0.409 | 0.13 | 3.138 | 0.006** |
| R Middle Frontal Cortex | 0.092 | 0.183 | 0.504 | 0.619 |
| L Transverse Temporal Cortex | 0.388 | 0.216 | 1.797 | 0.086 |

***6-7 hours of sleep***

| *Brain Node* | *B* | *SE* | *t* | *p* |
|---|---|---|---|---|
| L Cuneus | 0.467 | 0.244 | 1.913 | 0.061 |
| L Lateral Occipital Cortex | 0.5 | 0.226 | 2.2 | 0.032* |
| L Lingual Gyrus | 0.457 | 0.197 | 2.31 | 0.025* |
| R Paracentral Gyrus | 0.446 | 0.209 | 2.126 | 0.040* |
| L Pars Triangularis | 0.113 | 0.245 | 0.463 | 0.645 |
| R Middle Frontal Cortex | 0.017 | 0.191 | 0.089 | 0.93 |
| L Transverse Temporal Cortex | 0.396 | 0.238 | 1.664 | 0.104 |

***7-8 hours of sleep***

| *Brain Node* | *B* | *SE* | *t* | *p* |
|---|---|---|---|---|
| L Cuneus | 0.419 | 0.231 | 1.81 | 0.079 |
| L Lateral Occipital Cortex | 0.026 | 0.255 | 0.103 | 0.918 |
| L Lingual Gyrus | 0.282 | 0.221 | 1.293 | 0.205 |
| R Paracentral Gyrus | 0.555 | 0.215 | 2.574 | 0.014* |
| L Pars Triangularis | -0.2 | 0.207 | -0.968 | 0.34 |
| R Middle Frontal Cortex | 0.308 | 0.224 | 1.37 | 0.178 |
| L Transverse Temporal Cortex | 0.394 | 0.278 | 1.415 | 0.165 |

***Greater than 8 hours of sleep***

| *Brain Node* | *B* | *SE* | *t* | *p* |
|---|---|---|---|---|
| L Cuneus | 0.118 | 0.192 | 0.615 | 0.542 |
| L Lateral Occipital Cortex | 0.003 | 0.182 | 0.021 | 0.983 |
| L Lingual Gyrus | -0.026 | 0.143 | -0.186 | 0.854 |
| R Paracentral Gyrus | -0.094 | 0.187 | -0.506 | 0.616 |
| L Pars Triangularis | -0.138 | 0.149 | -0.926 | 0.362 |
| R Middle Frontal Cortex | 0.187 | 0.177 | 1.058 | 0.297 |
| L Transverse Temporal Cortex | -0.106 | 0.177 | -0.601 | 0.553 |

*Note: * < .05; ** <.01; *** <.001*